\documentclass[]{emulateapj}
\usepackage{apjfonts}
\newcommand{\be}{\begin{equation}}
\newcommand{\ee}{\end{equation}}
\newcommand{\bea}{\begin{eqnarray}}
\newcommand{\eea}{\end{eqnarray}}
\newcommand{\Msun}{M_{\odot}}

\def\kpc{\ {\rm kpc}}
\def\pc{\ {\rm pc}}
\def\yr{\ {\rm yr}^{-1}}
\def\sec{\ {\rm s}^{-1}}
\def\cm{\ {\rm cm}^{-3}}
\def\kms{{\ }{\rm km}\,{\rm s}^{-1}}
\newcommand{\comment}[1]{}

\shortauthors{CONROY \& SPERGEL}
\shorttitle{MULTIPLE STELLAR POPULATIONS IN GLOBULAR CLUSTERS}

\begin{document}
\journalinfo{The Astrophysical Journal}
\submitted{Submitted to the Astrophysical Journal}

\title{On the formation of multiple stellar populations in globular
  clusters}

\author{Charlie Conroy \& David N. Spergel}
\affil{Department of Astrophysical Sciences, Princeton
  University, Princeton, NJ 08544, USA}

\begin{abstract}

  Nearly all globular clusters (GCs) studied to date show evidence for
  multiple stellar populations, in stark contrast to the conventional
  view that GCs are a mono-metallic, coeval population of stars.  This
  generic feature must therefore emerge naturally within massive star
  cluster formation.  Building on earlier work, we propose a simple
  physical model for the early evolution (several $10^8$ yr) of GCs.
  We consider the effects of stellar mass-loss, type II and prompt
  type Ia supernovae, ram pressure, and accretion from the ambient
  interstellar medium (ISM) on the development of a young GC's own gas
  reservoir.  In our model, type II SNe from a first generation of
  star formation clears the GC of its initial gas reservoir.  Over the
  next several $10^8$ yr, mass lost from AGB stars and matter accreted
  from the ambient ISM collect at the center of the GC.  This material
  must remain quite cool ($T\sim10^2$K), but does not catastrophically
  cool on a crossing time because of the high Lyman-Werner flux
  density in young GCs.  The collection of gas within the GC must
  compete with ram pressure from the ambient ISM.  After several
  $10^8$ yr, the Lyman-Werner photon flux density drops by more than
  three orders of magnitude, allowing molecular hydrogen and then
  stars to form.  After this second generation of star formation, type
  II SNe from the second generation and then prompt type Ia SNe
  associated with the first generation maintain a gas-free GC, thereby
  ending the cycle of star formation events.  Our model makes clear
  predictions for the presence or absence of multiple stellar
  populations within GCs as a function of GC mass and formation
  environment.  While providing a natural explanation for the
  approximately equal number of first and second generation stars in
  GCs, substantial accretion from the ambient ISM may produce fewer
  chemically peculiar second generation stars than are observed.
  Analyzing intermediate-age LMC clusters, we find for the first time
  evidence for a mass threshold of $\sim10^4\Msun$ below which LMC
  clusters appear to be truly coeval.  This threshold mass is
  consistent with our predictions for the mass at which ram pressure
  is capable of clearing gas from clusters in the LMC at the present
  epoch.  Recently, claims have been made that multiple populations
  within GCs require that GCs form at the center of their own dark
  matter halos.  We argue that such a scenario is implausible.
  Observations of the young and intermediate-age clusters in the LMC
  and M31 will provide strong constraints on our proposed scenario.

\end{abstract}



\section{An observational puzzle}
\label{s:intro}

Globular clusters (GCs) have historically been considered coeval,
mono-metallic, gravitationally bound collections of stars.  In the
past several years, high precision photometric and spectroscopic
observations have led to a radical revision of this picture.

High resolution spectra of stars within nearly all GCs studied to date
reveal internal spreads in light element abundances such as C, N, Na,
O, Mg, and Al, beyond what can be explained by measurement errors
\citep[see review in][]{Gratton04}.  The magnitude of the internal
spread varies considerably from cluster to cluster, though there are
noticeable trends with cluster mass and orbital properties
\citep{Carretta06, Carretta10b}.  Intriguingly, internal spread in the
Fe-peak elements and the type II supernovae product Ca is limited to
only the most massive GCs \citep[$\omega$Cen, M22, M54, and Terzan
5;][]{Marino09, Carretta09a, Ferraro09, Carretta10a, Carretta10b}.
The most massive clusters also show indirect evidence for a large
spread in He abundances \citep{Gratton10}.  Abundance variations have
been detected in main sequence stars \citep{Gratton01}, indicating
that the observed variation arises from stars forming out of different
material, as opposed to being due to some unknown mixing process,
which could only occur along the giant branch.  While much attention
has been paid to internal abundance spreads recently, such internal
spreads have been known for over 30 years \citep{Cohen78, Kraft79,
  Peterson80, Freeman81, Smith82a, Smith83}.

Photometry from the {\it Hubble Space Telescope (HST)} has
demonstrated that many GCs contain multiple sub-giant branches and at
least two ($\omega$Cen and NGC 2808) contain multiple main sequences
\citep[see review in][]{Piotto09}.  Variations in the light element
abundances have been conclusively associated with the multiple
sequences observed in the color-magnitude diagram
\citep[CMD;][]{Yong08, Marino08, Carretta09b, Milone10}, demonstrating
that these two phenomena are intimately linked.  The population with
enhanced abundance patterns is more centrally concentrated than the
`normal' population in the GCs NGC 1851 \citep{Zoccali09} and
$\omega$Cen \citep{Sollima07}, and perhaps many others as well
\citep{Carretta09b}.

In the old Milky Way (MW) GCs the relative numbers of stars with
`anomalous' and normal abundances ratios is approximately equal, with
little dependence on metallicity \citep[e.g.,][]{Smith82a,
  Carretta09b}.  This fact imposes strong constraints on formation
scenarios, as we explain in later sections.  The relative numbers of
anomalous and normal stars in GCs in other galaxies (e.g., the LMC,
M31) is not known, but would provide new insight.

Additional insight has come from the strong observed correlations
between various light element abundances within individual GCs.  The
most striking is the anti-correlation between Na and O
\citep[e.g.,][]{Kraft93, Ivans99, Carretta09b}.  This correlation
arises naturally when material of standard (e.g., solar) abundance
ratios is mixed with material that has been
processed\footnote{Throughout the text we will use `processed' and
  `un-processed' to refer specifically to the nuclear processing of
  material at the temperatures required to explain the observed light
  element abundance variations, i.e., at $T>10^7$K.} at temperatures
$>10^7$K.  At such temperatures the Na-Ne and CNO cycles are active,
with the former producing Na and the latter depleting O.  At similar
temperatures the Mg-Al cycle is active, explaining observed
correlations between these elements as well.

The discovery of significant Li and F abundances in the second stellar
generation \citep[i.e., associated with stars of high Na and low O
abundances;][]{Pasquini05, Smith05} has provided yet another puzzle
and clue.  These elements are fragile, especially Li which burns at
$T\gtrsim10^6$K.  The existence of such fragile elements in the
atmospheres of second generation stars that also show strong O
depletion and Na enhancement suggests that these stars formed out of
at least two kinds of material; i.e., from matter exposed to
$T>10^7$K, and additional material that was never heated above
$T\sim10^6$K.  The interpretation of Li is however complicated by the
possibility that Li may, under special circumstances, actually be
produced within AGB stars \citep[see e.g.,][]{Ventura10}.

Multiple stellar populations have also been detected in
intermediate-age ($\sim1$ Gyr) and old LMC clusters \citep{Mackey08,
  Goudfroij09, Milone09, Mucciarelli09}, and possibly in the old GCs
within the Fornax dSph galaxy \citep{Letarte06}.  These observations
indicate that the multiple stellar population phenomenon is not
specific to the MW.  Observations of LMC clusters are of course
hampered by the much larger distance modulus to the LMC.  Despite this
limitation, the intermediate-age clusters offer a new window into the
internal age spreads because the main sequence turn-off point at these
ages is a strong function of time.  Small age differences are
therefore readily noticeable in the CMD.  Analysis of {\it HST}-based
CMDs have shown that the spread observed in the main sequence turn-off
of intermediate-age LMC clusters can be explained with an internal age
spread of a few $10^8$ yr.

The only class of star clusters known {\it not} to contain multiple
stellar populations are the open clusters in the MW \citep{deSilva09,
  Martell09}, which have typical masses of a few thousand solar
masses, although some, such as NGC 7789, have masses of
$\sim10^4\Msun$.

These observations have led to the unavoidable conclusion that the
majority of GCs studied to date harbor multiple stellar populations.
For all but the most massive ones, GCs are still considered chemically
homogeneous in Fe-peak and elements arising primarily from SNe type II
such as Ca.  The light element variations have been detected in both
young and old clusters, both metal-poor and metal-rich
\citep{Martell09}, and are noticeably absent in the open clusters.

From these observations the following timeline in the early evolution
of GCs has emerged.  Within GCs, a first generation of stars form.
Type II SNe then remove any remaining gas from the GC.  After the
epoch of type II SNe, mass from evolved stars is cycled through
temperatures of $T>10^7$K and then is returned to the gaseous
reservoir of the GC.  After a few $10^8$ yr, a second generation of
stars forms from a mix of processed and un-processed material.  Star
formation permanently ceases after the formation of the second
generation.  This process does not occur in open clusters, nor in the
field.  This basic timeline has, in one form or another, been
discussed by many authors \citep[e.g.,][]{Cottrell81, Smith87,
  Carretta10c}.

Any theory of GC formation must be embedded into our broader
cosmological theories of structure formation.  Many lines of evidence
suggest that galaxy formation is an hierarchical process where dark
matter halos serve as sites for assembling baryons and converting
baryons into stars.  One of the inevitable predictions of this
bottom-up scenario is that MW GCs likely formed in environments very
different from the $\sim200\kms$ dark matter dominated halo where they
now reside.  One possibility, which we discuss in $\S$\ref{s:dm}, is
that GCs form in the centers of small dark matter halos.  This
scenario would imply that isolated GC are embedded in extended dark
matter halos.  This paper emphasizes another possibility: GCs form
within small gas-rich dwarf galaxies --- the building blocks of the
present MW.  We emphasize that the MW is an evolving galactic system
so that estimates of ram pressure and gaseous accretion must consider
the likely environment in which a GC formed at $z\sim2-10$, rather
than its current location in the MW today.

We have outlined above only the most basic sketch of what must occur
to explain the observations.  In the next section we critically assess
previous, more detailed scenarios for the development of multiple
stellar populations within GCs.  Following this assessment, we
describe our own model for the early evolution of GC stellar
populations that includes several novel ingredients, and is, at least
in certain respects, more plausible than other scenarios.  We also
present an analysis of intermediate-age clusters in the LMC that
provides confirmation of a key aspect of our model.  We conclude by
commenting on various observations that may shed new light on this
exciting observational puzzle.

\section{Previous efforts}
\label{s:prev}

There are at least three major issues that require explanation before
any satisfactory theory for the formation of multiple populations in
GCs can be accepted.  These are: 1) understanding how the gaseous
reservoir within a GC can remain within the shallow potential well for
several $10^8$ yr; 2) identifying which stars are responsible for
processing material at $T>10^7$K; and 3) explaining how a second
generation of stars can form with a current total mass comparable to
the first generation.  In this section we will assess the plausibility
of previous efforts at addressing these outstanding issues.

\subsection{Can mass lost from first generation stars remain bound to
  the GC?}

A serious challenge to any theory for the formation of multiple
stellar populations is the shallow potential wells of GCs.  If gas is
to remain bound within GCs, then one of the following two scenarios
must occur: 1) the gas must both be ejected from stars below the GC
escape speed and must remain cool so that the internal dispersion is
$\lesssim10\kms$; or 2) the GC must be at the gravitational center of
a more massive system with the escape speed from the larger system
sufficiently large to retain stellar winds of a range of speeds.

An additional obstacle is ram pressure.  Many GCs cross the disk of
the MW at high velocity, which results in a strong ram pressure force
felt by gas within GCs.  Under certain conditions, ram pressure may
therefore prohibit the formation of a gaseous reservoir within GCs
\citep[see e.g.,][]{Frank76, Gnedin02}.

The shallow potential wells of GCs and their susceptibility to ram
pressure has led some to resurrect the notion that GCs form at the
center of their own dark matter halos at high redshift
\citep[e.g.,][]{Bekki06, Bekki07, Carretta10c}.  The feasibility of
this scenario is assessed in the following section.

\subsubsection{Formation of GCs at the centers of small dark matter
  halos?}
\label{s:dm}

Early theories regarding the formation of GCs included the possibility
that they form within extended dark matter halos at high redshift
\citep{Peebles84}.  This scenario fell out of favor following the
observation of thin tidal tails surrounding many GCs
\citep[e.g.,][]{Grillmair95, Odenkirchen03}, because numerical
simulations showed that such tidal tails do not form if GCs reside
within extended halos \citep{Moore96b}.

The formation of GCs at the center of their own dark matter halos has
received renewed interest thanks to numerical work that has shown that
extended halos of GCs could be tidally stripped away by the present
epoch \citep{Bromm02, Mashchenko05b}.  In this picture, the constraint
from tidal features demonstrates only that GCs at the present epoch
are not embedded within extended massive halos, but offers no insight
into the formation environment of GCs.

This scenario has been adopted in order to explain the formation of
multiple stellar populations within GCs \citep[e.g.,][]{Freeman93,
  Bekki06, Bekki07, Boker08, Carretta10c} for the principle reason
that GCs embedded within massive halos do not experience ram pressure
and may also easily retain stellar winds.  In our opinion, the
motivation for dark matter halos from ram pressure arguments is in
error.  Recent arguments regarding ram pressure have been based
directly on ram pressure calculations for GCs orbiting the MW {\it at
  the present epoch}.  Indeed, no GCs today show evidence of a gaseous
reservoir, and ram pressure is thought to be responsible \citep[see
e.g.,][]{Tayler75, Frank76}.  These calculations assume mass-loss
rates appropriate for $\sim10^{10}$ yr old populations, and orbital
velocities characteristic of the present MW.  In contrast, at the
epoch relevant for the formation of multiple populations (e.g., the
formation epoch of GCs), the characteristic orbital velocities were
almost certainly substantially smaller, probably by an order of
magnitude, and the mass-loss rates were an order of magnitude larger.
We will demonstrate quantitatively in a later section that these two
effects combine to significantly reduce the efficiency of ram pressure
so that all MW and LMC clusters that show evidence for multiple
stellar populations were likely impervious to this effect in their
formation environment, without appeal to dark matter halos.  In short,
ram pressure is {\it not} an important physical process for the mass
ranges and formation environments of the current set of clusters with
known multiple populations.

There are additional arguments against the formation of GCs within
dark matter halos.  The most striking is the observation of multiple
populations within intermediate-age LMC clusters.  These clusters
surely are not embedded within massive halos, and yet they clearly are
capable of forming multiple populations.  If these clusters can form
multiple populations without being embedded in a massive halo, why not
MW GCs?

If GCs formed at the center of their own halos, then the existence of
GCs within dwarf spheroidals would be extremely difficult to
understand because they would have very short dynamical friction
times.  For example, Fornax contains five GCs.  In the absence of an
extended massive halo, the friction timescales are already
uncomfortably short if Fornax resides within a standard dark matter
halo \citep[several Gyr;][]{Goerdt06}.  If these GCs were embedded in
massive halos, their dynamical friction timescales would be $\ll1$ Gyr
since friction scales with the total GC mass.

A final argument against GC formation within extended halos is the
structure and kinematics of very isolated GCs both in the MW and M31.
NGC 2419 is a remote MW GC at a distance of $\approx90$ kpc from the
Galactic center, has a mass of $\sim10^6\Msun$, and half-mass radius
of $\approx20$ pc.  Based on radial velocity measurements of stars
within NGC 2419, \citet{Baumgardt09} measured a mass-to-light ratio
that was consistent with a pure stellar population.  It is hard to
imagine how tidal stripping could effect this cluster, and so the lack
of any evidence for a massive dark matter halo for NGC 2419 argues
against GCs generically forming within massive halos.  Recently, an
extremely isolated GC has been found within M31, at a distance of
$\approx200$ kpc from the center of M31 \citep{Mackey10}.  This
object, far removed from any strong tidal fields, should be another
ideal candidate to search for evidence of dark matter within GCs.

The formation of GCs within massive halos does however provide one
significant advantage over models that do not invoke formation within
halos.  At high redshift (e.g., $z=10$), the gas accretion rate onto a
$10\kms$ halo from the intergalactic medium is of the order $\rho_{\rm
  IGM}V4\pi R^2\approx6\times10^{-3}\Msun\yr$.  The accretion rates
are much higher at $z=10$ than the present epoch because the density
of the IGM is higher by a factor of $(1+z)^3=10^3$.  After several
$10^8$ yr, this would result in a substantial accumulation of gas,
which could help explain the relative numbers of first and second
generation stars (see $\S$\ref{s:nums}) and the high abundances of Li
and F.

In this section we have argued that 1) GCs need not form at the
centers of their own halos in order to survive ram pressure stripping
and 2) regardless of ram pressure considerations, not all GCs could
have formed within their own halos given both dynamical constraints
and observations of intermediate-age GCs in the LMC and M31.
Nonetheless, we emphasize the possibility that {\it some} GCs formed
at the centers of their own halos.  Nuclear star clusters reside at
the centers of galaxies, are massive ($\sim10^6-10^8\Msun$), compact
(half-light radii of a few pc), and span a wide range of ages
\citep[$10^7-10^{10}$ yr;][]{Boker04, Walcher05, Walcher06}.  The
youngest nuclear star clusters must certainly have formed where they
are now observed.  The formation environment of the older clusters is
less clear because they may have migrated toward the center via
dynamical friction.  In any event, the existence of nuclear star
clusters at least suggests that GC-like objects can in some cases form
at the centers of their own dark matter halos.  As we discuss below,
the most massive GCs in the MW may be examples of such objects.

\subsection{Identifying the stellar type responsible for processing
  material to $T>10^7$K}
\label{s:stype}

Based on nucleosynthetic constraints, there are two plausible
candidate donors of processed material: massive ($\gtrsim 20\Msun$)
rotating stars \citep{Decressin07}, and massive ($\approx4-8\Msun$)
AGB stars \citep{Ventura01, DAntona02}.  Unfortunately, both of these
proposed sites suffer from serious theoretical uncertainties (e.g.,
compare the AGB models of \citet{Karakas07} and \citet{Ventura08}).
The relevant cross-sections for the production of Na are uncertain by
a factor of $\sim1000$ \citep[see discussion in][]{Ventura08}, making
detailed comparisons between models and data difficult.  For example,
\citet{Ventura08b} were able to construct AGB models whose ejecta
abundance patterns were consistent with the abundances of observed
second generation stars only after they modified several key yet
uncertain nuclear cross sections.  Models for massive rotating stars
suffer from significant uncertainties not only in the mass loss rates
but also in the efficiency of meridional circulations to mix processed
material to the envelope.

The massive rotating stars model suffers from several shortcomings.
First, the timescale for these stars to lose mass is comparable to the
timescale for type II SNe.  It is difficult to imagine how new stars
could form in an environment where massive stars are constantly
exploding.  Even if this could occur, the second generation would then
show significant over abundances in $\alpha$-elements compared to the
first generation, but this is not observed \citep[see also][who make a
similar point]{Renzini08}.

Moreover, the short timescales involved would imply that a different
physical mechanism is at work in the intermediate-age LMC clusters,
where CMD studies have shown internal age spreads of order $10^8$ yr
\citep{Milone09}.  In contrast, the AGB stars evolve on timescales
similar to the observed age spread in LMC clusters, providing a
natural explanation for the observed spread.

Finally, the short evolutionary timescales for the massive rotating
stars makes it difficult to explain the observed Li and F abundances.
The most plausible explanation for the relatively high abundances of
these elements is that the second generation formed from a mix of
un-processed and processed material.  As we discuss in later sections,
it is very difficult to accrete a sufficient amount of un-processed
material in only $\sim10^7$ yr.

Recently, \citet{deMink09} proposed a scenario whereby massive
binaries shed a large fraction of their mass due to non-conservative
mass and angular momentum transfer.  We believe that this scenario
suffers from many of the shortcomings of the winds from massive stars
scenario, as outlined above, and is therefore also disfavored.

We instead favor AGB stars as the source of processed material out of
which the second generation stars form.  Besides the fact that they
are the only remaining option, the evolutionary timescale of massive
AGB stars is comparable to the inferred internal age spread within
intermediate age GCs in the LMC.

\subsection{Constraints imposed by the relative numbers of first and
  second generation stars}
\label{s:nums}

As emphasized by many authors, it is very difficult to understand the
observational fact that within most GCs the number of first and second
generation stars is approximately equal.  For example, assuming a
standard \citet{Kroupa01} IMF, and assuming that AGB stars shed all of
their mass aside from a $0.6\Msun$ white dwarf remnant mass, only
$\approx9$\% of the total initial stellar mass of the system can be
returned to the surrounding GC from AGB winds with initial masses of
$4-8\Msun$\footnote{Our returned mass fraction is substantially
  smaller than the fractions quoted in \citet{Bekki06} because they
  assume AGB stars down to $1\Msun$ can contribute to the GC's gaseous
  reservoir.  The main sequence lifetime of such stars is
  substantially longer than $\sim10^8$ yr, and in addition it is
  believed that hot bottom burning, which is the source of many of the
  abundance anomalies, does not occur in such low mass stars.  We
  therefore believe our fraction, which is based on the expected AGB
  mass range for hot bottom burning, to be more accurate.  Expanding
  the adopted mass range to $3-8\Msun$ results in a mass return
  fraction only slightly larger ($13$\%) than what we quote for the
  $4-8\Msun$ range.}.  If 100\% of the AGB ejecta were converted into
stars, the population ratio should be approximately ten first
generation stars for every one second generation star.

Several scenarios have been proposed to account for the discrepancy
between the above expectation and the observed ratio.  Early proposals
focused on ad hoc variations to the IMF in order to produce the
observed relative numbers \citep[e.g.,][]{Smith82b}.  Recent work has
even attempted to {\it constrain} the IMF of the first generation in
this way \citep{DAntona04, Prantzos06}.

More recently, \citet{DErcole08} proposed a solution that required the
first generation to have been substantially more massive (by a factor
of $10-100$) than its present mass.  If the first generation was born
with mass segregation already in place, then stellar evolutionary and
dynamical effects could result in the loss (i.e., unbinding) of a
significant number of first generation stars.  The second generation,
being born with higher concentration, would be relatively immune to
such a process, and so the majority of second generation stars would
remain bound.  The preferential loss of first generation stars could
then result in comparable numbers of first and second generation stars
at late times.

The proposal that second generation stars form purely from AGB ejecta
suffers from several shortcomings, including the required level of
fine-tuning.  If we assume that the first generation donates 10\% of
its mass to the GC gas reservoir via AGB winds and a star formation
efficiency of 30\%, the resulting total mass in the first and second
generations will be equal if the first generation losses 97\% of its
initial mass.  This assumes that no stars from the second generation
are lost; any mass lost from the second generation would require an
even larger fraction of mass-loss from the first generation. In the
\citet{DErcole08} scenario, the first generation of every GC must lose
$\gtrsim97$\% of its mass so that the observed population ratios are
always of order unity. Without a natural mechanism to explain why the
first generation always sheds enough mass so that its final mass is
comparable to the second generation, the mass lost must be fine tuned
{\it ad hoc}.  One would also expect isolated GCs such as NGC 2419
that experience weak tidal fields to contain many more (by factors of
$10-100$) first generation compared to second generation stars.
Abundance measurements of stars within NGC 2419 currently do not exist
but would clearly shed light on this issue.

A second concern with the D'Ercole et al. solution is the large
remnant population left behind by the disproportionately large
fraction of massive first generation stars at the center of the GC.
In this scenario, a GC with present mass $10^5\Msun$ and equal numbers
of first and second generation stars would require an initial stellar
mass of $\sim10^7\Msun$ first generation stars in order to generate
enough gas from AGB winds to produce a second generation with total
mass $\sim10^5\Msun$.  For a standard \citet{Kroupa01} IMF,
approximately 10\% of the total initial mass is constituted by
$M>8\Msun$ stars.  If 10\% of these massive first generation stars are
concentrated toward the center because of mass segregation and then
explode, a population of neutron stars and black holes will be left
behind, and remain bound, with a total mass of $\sim10^4\Msun$!

Such a large population of massive remnants would not only be
detectable as an additional dark mass component concentrated toward
the dynamical center, but would also imply a very high incidence of
low-mass X-ray binaries (LMXB).  Neither of these predictions are
supported by the observations \citep[e.g.,][]{vanderMarel04,
  Verbunt87}, although there are notable caveats.  The observed
frequency of LMXBs can be naturally explained by the high rate of
binary formation in GCs compared to the field, although this mechanism
is quite uncertain \citep{Fabian75, Verbunt87}.  We have assumed that
a significant fraction of neutron stars remain bound to the cluster,
which may not be true if the majority of GC neutron stars are born
with high kick velocities \citep[as appears to be the case for pulsars
in the MW;][]{Hansen97}.  Appealing to IMF variations to explain
the relative numbers of first and second generation stars would also
produce significantly more neutron stars and black holes, as pointed
out by \citet{DAntona04}.  A more careful analysis of the expected
remnant population in GCs may yield stronger constraints on these
models.

Recently, \citet{Sills10} have investigated the importance of stellar
collisions in boosting the number of fast rotating massive stars and
intermediate mass stars (i.e., AGB stars) in GCs.  These authors find
that stellar collisions are not sufficiently numerous for their
products to help explain the relative number of first and second
generation stars.

In the following section we propose an alternative scenario for
resolving this discrepancy that is based on adding a significant
amount of gas via accretion from the ambient ISM.


\section{A plausible model for the formation of multiple stellar
  populations in globular clusters}

In the previous section we evaluated a number of current scenarios
that address aspects of the formation of multiple stellar populations
within GCs.  In this section we propose an alternative scenario for
the formation of multiple stellar populations in GCs.

\subsection{Model overview}

Our scenario can be summarized as follows.  A first generation of
stars forms out of gas that has already been pre-enriched to GC
abundances.  Type II SNe expel the remaining gas and thus shut off
star formation.  Over several $10^8$ yr, mass lost from massive AGB
stars is returned to the gaseous reservoir of the GC.  In addition,
{\it over this same time period, mass is accreted onto the GC from the
  ambient ISM}.  This un-processed gas is {\it incompletely} mixed
with the AGB ejecta.  All of the gas within the GC must be relatively
cold ($T\sim10^2$K), which implies efficient cooling.  However, this
gas does not cool catastrophically for several $10^8$ yr because of
the high Lyman-Werner photon density, which photodissociates molecular
hydrogen (H$_2$).  After several $10^8$ yr, the Lyman-Werner photon
density drops by more than three orders of magnitude (see Figure
\ref{fig:lwp}), H$_2$ rapidly forms, and star formation is triggered
within the GC's gas reservoir.  A second generation of stars is
born.  Type II SNe from the second generation clear out the remaining
gas\footnote{Of course, star formation must occur on a short enough
  timescale so that type II SNe do not prohibit the formation of the
  second generation.  Star formation must also occur over a short
  timescale during the formation of the first generation.  This has
  been a long-standing theoretical problem in the formation of bound
  GCs.}.  On a similar timescale as the formation of the second
generation (several $10^8$ yr), prompt type Ia SNe from the first
generation begin to explode.  These SNe then act to maintain a
gas-free GC environment, thereby permanently ending star formation.

Notice that GCs do {\it not} form in the centers of dark matter halos
in our scenario.  Rather, in our scenario, GCs form wherever ISM
conditions are favorable, including within gas-rich dwarf galaxies and
interacting systems; the former being the presumed building blocks of
the present day MW.  Evidence for efficient GC formation within
gas-rich dwarfs can be found in the high specific frequencies of GCs
within dwarfs compared to $L^\ast$ galaxies \citep{Lotz04}.

This model contains several novel ingredients with respect to previous
work.  First, we invoke significant accretion from the ambient ISM
during the development of the GC gas supply \citep[see
also][]{Pflamm-Altenburg09, DErcole10b}.  Second, we appeal to the
importance of the high Lyman-Werner photon flux density in delaying
star formation for several $10^8$ yr.  Third, we assert (and quantify
below) that GC formation in the centers of dark matter halos is not a
necessary condition for the formation of multiple stellar populations
because ram pressure is not effective at stripping the GC's gas
reservoir at early times.

We emphasize that certain aspects of this model, described in detail
below, are not able to naturally explain the properties of the most
massive MW GCs (e.g., $\omega$Cen, NGC 2808, M22, and M54), which show
evidence for multiple {\it distinct} stellar populations, and appear
to require very high He abundances ($Y\sim0.4$).  We consider these
objects separately in $\S$\ref{s:res}.

We now provide observational and theoretical motivations for each of
these elements.

\begin{itemize}

\item As summarized in $\S$\ref{s:intro}, there is no internal spread
  in either the Fe-peak elements or the $\alpha$-elements (that are
  unambiguously associated with Type II SNe, e.g., Ca), except for the
  most massive GCs.  This implies that the second generation formed
  out of material that was {\it not} enriched by the type II SNe
  associated with the first generation, nor by Type Ia.  We conclude
  from this that type II SNe efficiently evacuated the remaining gas
  from the GC after the formation of the first generation.  Energetic
  arguments support this conclusion \citep[e.g.,][]{Dopita86,
    Baumgardt08}.  Notice that any ambient ISM that is later accreted
  onto the GC must also be unpolluted by these type II SNe.  It might
  be difficult to avoid this if GCs form at the center of their own
  dark matter halos.

\item Consideration of the effects of type II SNe leads us to disfavor
  winds from massive rotating stars as the source of the processed
  material \citep[i.e., the scenario proposed by][]{Decressin07}.
  These stars evolve on timescales comparable to the onset of type II
  SNe.  It is therefore very difficult to see how a second generation
  of stars could form out of the mass lost via winds from these
  massive stars both because the SNe would tend to expel the gas, and
  also because if the gas remained, it would be enriched in
  $\alpha-$elements, which is not observed \citep[see also
  $\S$\ref{s:stype} and][]{Renzini08}.  We therefore favor AGB ejecta
  as the source of the processed material.

\item Evidence for accretion of pristine material comes from the
  observed abundance variations and abundance correlations within GCs.
  \citet{Prantzos07}, \citet{Ventura08b} and \citet{Ventura09} have
  argued that the Na--O anti-correlation {\it requires} dilution from
  pristine material.  In addition, \citet{Pasquini05} and
  \citet{Prantzos07} have argued that the discovery of high abundances
  of Li and F in the second stellar generation requires significant
  amounts of pristine material.  Their argument is based on the fact
  that these elements are fragile; Li for example burns at
  $\sim2\times10^6$ K.  If the second generation formed purely from
  AGB ejecta, one might expect these stars to be free of Li and F.
  \citet{Ventura10} point out that Li can in fact be made in the
  convective envelopes of massive AGB stars via the Cameron-Fowler
  mechanism.  The implied yields are however strongly dependent on
  metallicity, stellar mass, and the adopted mass-loss rates. 

  The need for dilution from pristine material makes it easier to form
  a second generation of stars with total mass comparable to the first
  generation, since second generation stars apparently do not form
  {\it exclusively} out of AGB ejecta.  Notice that this un-processed
  material cannot have significantly different Fe-peak nor
  $\alpha$-element abundances, which has a number of implications.
  One implication is that the type II SNe that presumably cleaned out
  the GC after the first generation was born cannot have mixed with
  this un-processed material, at least for the MW GCs, for which
  constraining data are available.

\begin{figure}[!t]
\resizebox{3.6in}{!}{\includegraphics{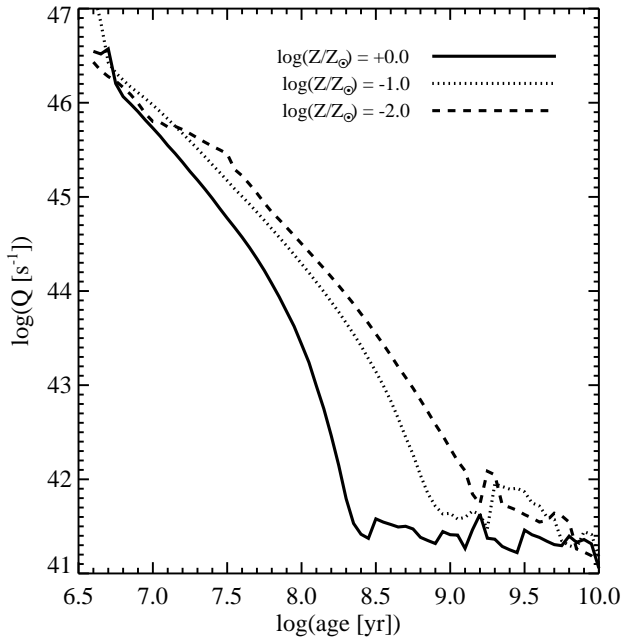}}
\caption{Evolution of the Lyman-Werner photon production rate for
  coeval stellar populations with metallicities of log$(Z/Z_\Sol)=0.0,
  -1.0$ and $-2.0$.  Rates are based on the stellar population
  synthesis models of \citet{Conroy09a} and are normalized such that
  one solar mass of stars formed instantly at $t=0.0$.  Notice the
  rapid decline in the rate at several $10^8$ yr.  Lyman-Werner
  photons are capable of destroying H$_2$ and so the precipitous drop
  in the photon production rate at $t>10^8$ yr implies favorable
  conditions for H$_2$ formation.}
\label{fig:lwp}
\end{figure}

\item Stars within GCs with abundance variations display a {\it range}
  of Na and O abundances, rather than simply two values as might be
  expected if the second generation formed from a chemically uniform
  gas reservoir.  We interpret this result as evidence of incomplete
  mixing in the gas between the processed and accreted ambient ISM
  material.  For example, \citet{Prantzos07} demonstrate that the full
  range of F and Li abundances can be reproduced in NGC 6742 and M4
  with a `dilution' factor of ambient ISM material ranging from 0.1 to
  250.  \citet{Ventura08b} and \citet{Ventura09} have shown that the
  full range of the Na--O anti-correlation can be explained if stars
  form from a mixture of AGB ejecta and pristine material.  Clearly,
  the GC gas reservoir must be quite heterogeneous prior to the second
  generation of star formation.

\item Over the past few years, there has been a growing recognition
  that most type Ia SNe are prompt rather than delayed explosions with
  timescales of Gyrs \citep{Scannapieco05}.  The prompt Ia SNe rate
  has been estimated to be roughly $10^{-12}$ SNe $\Msun^{-1}\,\yr$,
  which translates into $\approx10$ SNe per $10^5\Msun$ per $10^8$ yr
  \citep[e.g.,][]{Brandt10, Maoz10}, or $\sim10^{52}$ erg of energy
  per $10^5\Msun$ per $10^8$ yr.  The binding energy of $10^5\Msun$ of
  gas with radius 1 pc --- an extreme case --- is $\approx10^{51}$
  erg.  Once they begin to occur (i.e., after the delay time), prompt
  SNe Ia will therefore be sufficient to keep the GC gas-free.  This
  is an important element.  If there were no prompt Ia SNe, it is not
  clear why a third, fourth, etc. generation of stars would not form
  from the AGB ejecta of earlier generations.  The importance of
  prompt Ia SNe has also been emphasized by \citet{DErcole08}.

\item Young GCs produce high flux densities of Lyman-Werner photons
  ($912<\lambda<1100$\AA).  These photons are absorbed by H$_2$ and
  $\approx$16\% will lead to dissociation of the molecule.  In Figure
  \ref{fig:lwp} we show the flux density of Lyman-Werner photons as a
  function of time for a coeval stellar population.  Notice the
  precipitous drop in the Lyman-Werner flux density at several $10^8$
  yr.  We are interested in knowing if the flux density at $<10^8$ yr
  is sufficiently high to prohibit the formation of H$_2$ in the bulk
  of the gas within GCs.  This physical situation is analogous to
  photodissociation regions (PDRs) seen within the MW.  In our case we
  have many, possibly overlapping PDRs because stars are randomly
  spread throughout the gas within young GCs.  The following
  calculation is an order-of-magnitude estimate that proves to be
  sufficient for our purposes.

  We can compute the radius at which photodissociation of H$_2$ is
  equal to the formation of H$_2$ in a manner analogous to a
  Str\"{o}mgren sphere calculation.  At the temperatures and
  metallicities of interest ($T\approx100$K, log($Z/Z_\Sol)>-3$),
  H$_2$ forms most efficiently on dust grains, with a rate
  coefficient\footnote{The H$_2$ formation rate and its dependence on
    physical parameters is quite uncertain.  We adopt a simple linear
    scaling with metallicity because of the expected linear scaling
    between metallicity and dust mass fraction.  This scaling is
    supported by \citet{Tumlinson02} who find that the H$_2$ formation
    rate in the LMC and SMC is a factor of $\sim10$ lower than in the
    MW.} of $3\times10^{-17}\frac{Z}{Z_\Sol}$ cm$^3\,\sec$
  \citep{Jura75, Wolfire08}.  The balance between formation and
  destruction occurs at a radius:
  \noindent
  \be
  R_{\rm LW} = 0.15\pc \bigg(\frac{Q}{10^{46}\sec}\bigg)^{1/3}\,  
  \bigg(\frac{10^5\cm}{n}\bigg)^{1/3}\, \bigg(\frac{0.1Z_\Sol}{Z}\bigg)^{1/3},
  \ee
  \noindent
  where $Q$ is the photon production rate provided by starlight over
  the interval $912<\lambda<1100$\AA, $n$ is the gas density, and $Z$
  is the metallicity of the gas.  The value of $Q=10^{46}\sec$ is
  appropriate for a stellar age of $\approx10^7$ yr (see Figure
  \ref{fig:lwp}).  Notice that we have effectively assigned each GC
  star an average spectrum appropriate for a coeval population.

  This radius is to be compared to the mean interstellar spacing:
  \noindent
  \be
  l = 0.05\pc \bigg(\frac{10^6}{N_\ast}\bigg)^{1/3}
  \bigg(\frac{R'}{3\pc}\bigg),
  \ee
  \noindent
  where $N_\ast$ is the number of stars within a radius $R'$.  The gas
  within a young GC will form a considerable amount of H$_2$ only when
  $R_{\rm LW}<l$.  Based on the evolution in the flux density shown in
  Figure \ref{fig:lwp}, this will occur only after several $10^8$ yr,
  when $Q$ decreases by a factor of $>10^3$.  Notice also that even if
  90\% of the Lyman-Werner photons are absorbed by dust (so that the
  effective $Q$ is reduced by a factor of 10), the photon flux will
  still be sufficient to prevent significant formation of H$_2$ during
  the early evolution of GCs.  The drop in the Lyman-Werner flux
  density after several $10^8$ yr is so dramatic that our conclusion
  regarding H$_2$ photodissociation is not sensitive to details.

\item The $\sim10^8$ yr separation between first and second
  generations is the natural timescale for four completely separate
  but potentially relevant processes: the orbital time within galactic
  environments, the main sequence lifetime of massive
  ($\approx4-8\Msun$) AGB stars, the timescale for the onset of prompt
  SNe Ia \citep[e.g.,][]{Brandt10, Maoz10}, and the time when
  Lyman-Werner photon production drops precipitously for a coeval
  stellar population.  The temporal separation between first and
  second generations cannot be significantly longer than $10^8$ yr
  because otherwise prompt SNe Ia would clear out the GC gas, which
  would prevent the formation of the second generation.  The lack of
  an internal spread in Fe-peak elements provides additional support
  to this notion.  The orbital timescale may be relevant because tidal
  perturbations could be a trigger that sets off the second generation
  of star formation.  Once Lyman-Werner photon production drops, H$_2$
  can form, and the gas will catastrophically cool and, presumably,
  form stars.

\end{itemize}

\begin{figure*}[!t]
\begin{center}
\vspace{.5cm}
\resizebox{6.2in}{!}{\includegraphics{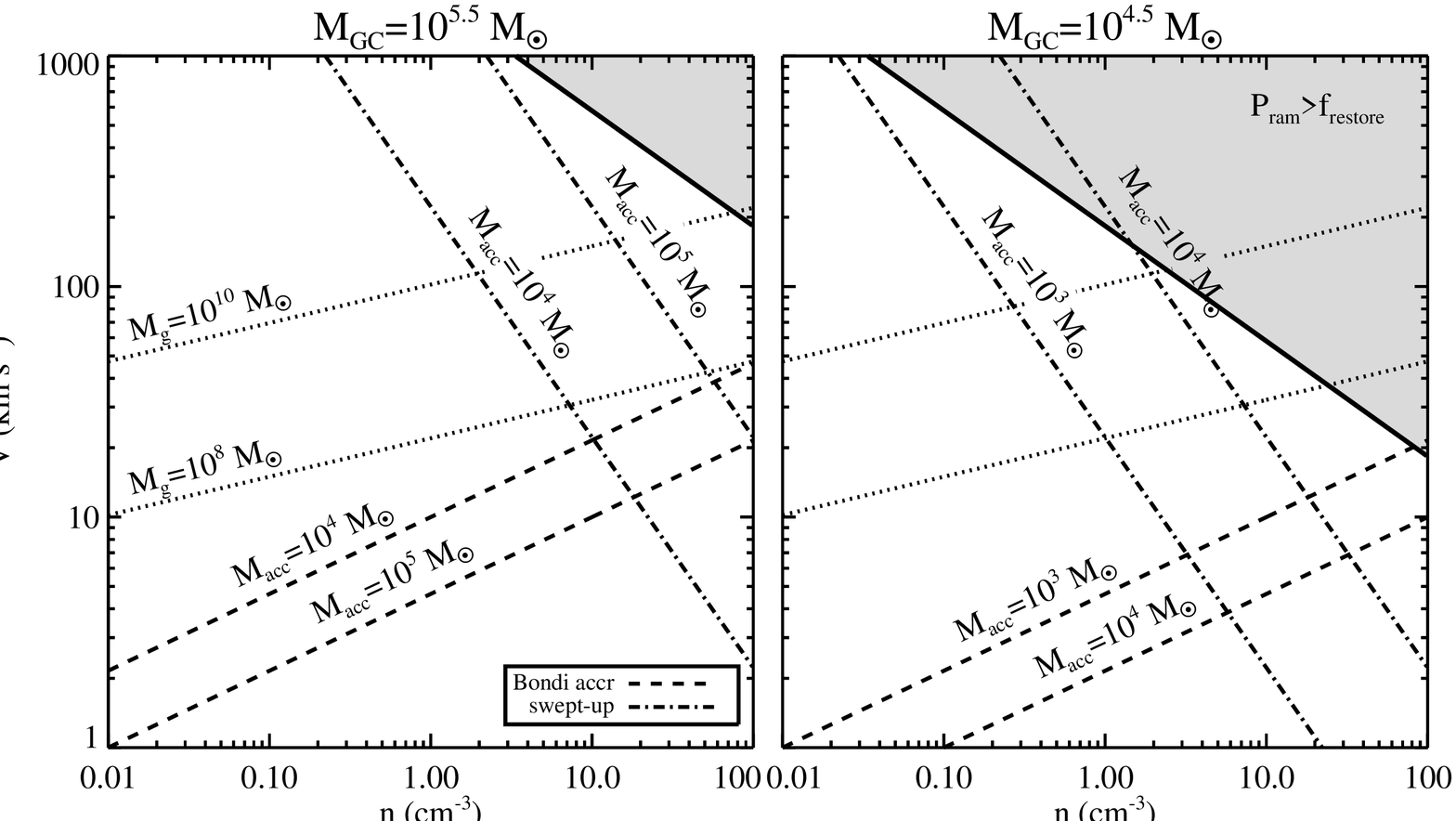}}
\end{center}
\vspace{0.5cm}
\caption{Comparison of the importance of ram pressure, Bondi
  accretion, and ambient ISM sweeping to the development of the gas
  reservoir within GCs (see Equations \ref{e:bondi}, \ref{e:ism}, and
  \ref{e:pram}).  In this model, GCs begin with a gas reservoir with
  mass equal to 10\% of the total mass (i.e., $f=0.1$), have a
  half-mass radius of 3 pc, and move with velocity $V$ through an
  ambient ISM with average density $n$.  Left panel shows results for
  a GC with mass $10^{5.5}\Msun$; right panel is for a GC of mass
  $10^{4.5}\Msun$.  Shaded zones are regions of parameter space where
  ram pressure is strong enough to remove the gas from a GC.  Lines
  show the amount of material accreted by the GC after $10^8$ yr by
  either Bondi accretion ({\it dashed lines}) or ambient ISM sweeping
  ({\it dot-dashed lines}).  We also show in this figure the relation
  between $V$ and $n$ for the systems within which the GCs may be
  orbiting ({\it dotted lines}).  This last relation is constructed by
  assuming gas-dominated self-gravitating systems (i.e., assuming
  $V^2=GM_g/R$) with gas masses indicated in the figure.}
\label{fig:main}
\vspace{0.5cm}
\end{figure*}

\begin{figure*}[!t]
\begin{center}
\vspace{.5cm}
\resizebox{6.2in}{!}{\includegraphics{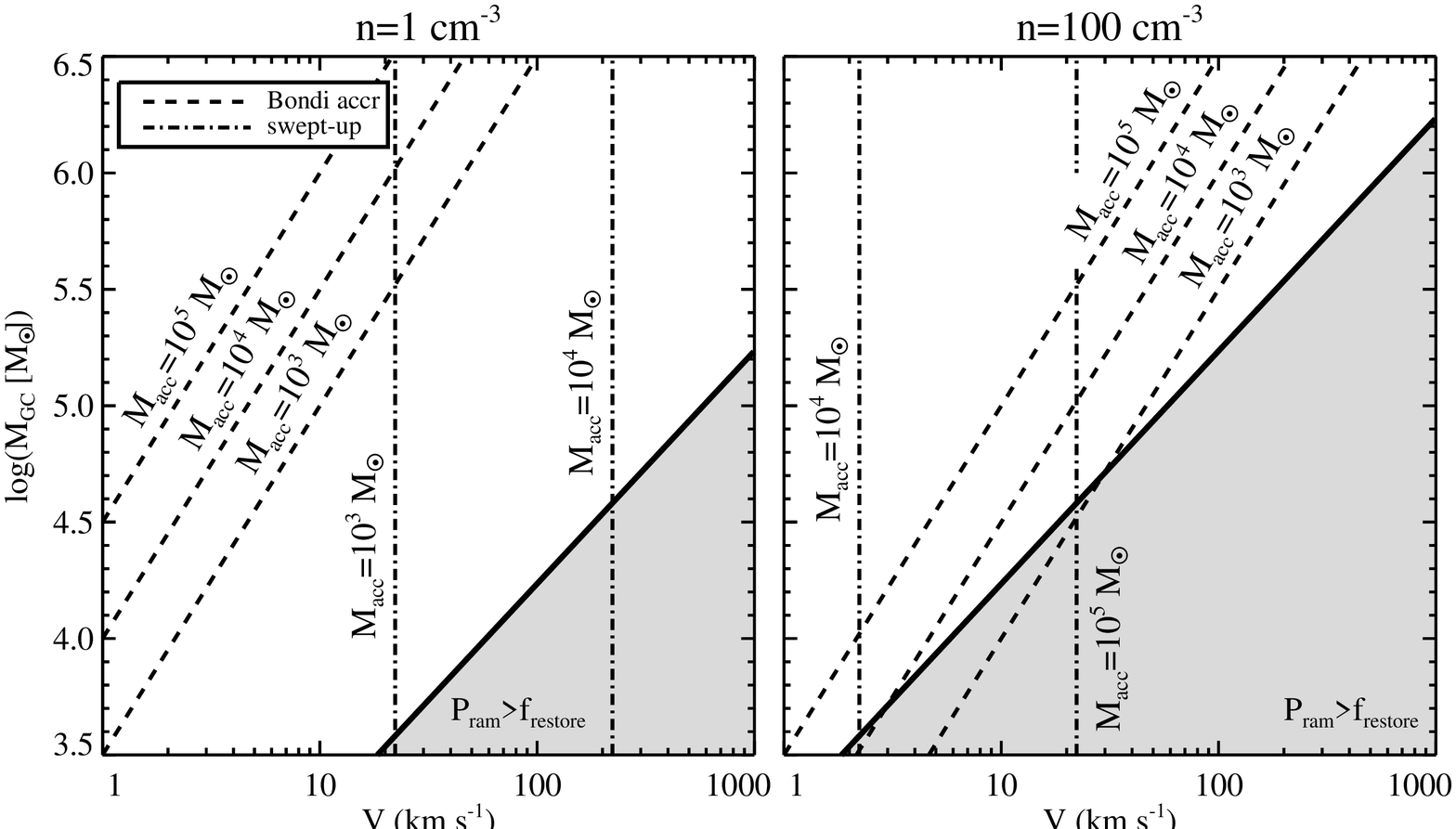}}
\end{center}
\vspace{0.5cm}
\caption{Comparison of the importance of ram pressure, Bondi
  accretion, and ambient ISM sweeping to the development of the gas
  reservoir within GCs.  This figure is similar to Figure
  \ref{fig:main} except for a fixed ambient density of $n=1\cm$ ({\it
    left panel}) and $n=100\cm$ ({\it right panel}), and a range of GC
  masses, $M_{\rm GC}$.  Lines and shaded regions are as in Figure
  \ref{fig:main}.}
\label{fig:mass}
\vspace{0.5cm}
\end{figure*}

In this section we have presented basic arguments favoring our adopted
scenario.  In the following section we expand upon the topic of the
development of a GCs gas supply.

\subsection{GC gas reservoir growth and retention}
\label{s:quant}

We will now explore the relative importance of ram pressure, Bondi
accretion, and the sweeping up of the ambient ISM via simple geometric
cross section, as a function of GC mass, $M$, relative velocity, $V$,
between the ambient ISM and the GC, and density of the ambient medium,
$n$, through with the GC is moving.  In the following discussion we
will consider Bondi accretion onto the cluster as a whole, not the
accretion of material onto individual stars.  Our primary objectives
in this section are twofold: to determine to what extent ram pressure
prohibits the formation of a second stellar generation, and to
determine whether accretion from the ambient ISM can provide a
significant source of gaseous material for a second generation of star
formation.

An object with mass $M$ embedded within an ambient medium with density
$n$ will accrete matter at the \citet{Bondi52} rate:
\noindent
\be
\dot{m_B} \approx 10^{-5}\,
\bigg(\frac{M}{10^5\Msun}\bigg)^2\bigg(\frac{n}{\cm}\bigg)
  \bigg(\frac{V}{10 \kms}\bigg)^{-3}\, \Msun \,{\rm yr}^{-1},
\ee
\noindent
where $V$ can be interpreted as the sum in quadrature of the sound
speed of the ambient medium and the velocity of the object through the
medium.  The accretion rate depends somewhat on the equation of state
of the ambient medium.

Over $10^8$ yr, Bondi accretion will result in a total mass accretion
of $M_B$.  After $10^8$ yr we therefore have:
\noindent
\be
\label{e:bondi}
\bigg(\frac{M_B}{10^3\Msun}\bigg) \approx 
\bigg(\frac{M}{10^5\Msun}\bigg)^2\bigg(\frac{n}{\cm}\bigg)
\bigg(\frac{V}{10 \kms}\bigg)^{-3},
\ee
\noindent
which, for a given $M_B$ and $M$ results in a simple relation
between $n$ and $V$.  In this equation $M$ represents the total mass
within the GC.  Below, we will approximate $M$ by the ``initial'' GC
mass (i.e., the mass before any gas is accreted).  This approximation
yields a lower limit to the total mass accumulated via Bondi accretion. 

In addition to Bondi accretion, an object moving with velocity $V$
through an ambient medium with density $n$ and cross section $A$ will
sweep up material at a rate given by $\dot{m_S} = \rho V A$.  The
scenario we imagine here is one where a seed gas reservoir is already
present within the GC.  This seed will then, by collisional forces, be
able to sweep up additional gaseous material.  This process will occur
so long as the ram pressure is less than the restoring force provided
by the potential well of the GC (see below).  This ambient ISM
sweeping results in a mass accretion rate of:
\noindent
\be
\dot{m_S} \approx 1.5\times10^{-5}\bigg(\frac{n}{{\rm
    cm}^{-3}}\bigg)\bigg(\frac{V}{10^2 \kms}\bigg)
\bigg(\frac{R}{3\pc}\bigg)^2\, \Msun \,{\rm yr}^{-1},
\ee
\noindent
where $R$ is the characteristic radius of the gas reservoir within the
GC.  Notice the relatively strong dependence on radius.  After $10^8$
yr, a total mass of $M_S$ will be swept-up, and the following relation
will hold:
\noindent
\be
\label{e:ism}
\bigg(\frac{M_S}{1.5\times10^3\Msun}\bigg) \approx \bigg(\frac{n}{\cm}\bigg)\bigg(\frac{V}{10^2\kms}\bigg)\bigg(\frac{R}{3\pc}\bigg)^2,
\ee
which, for a given $M_S$ and $R$ results in a simple relation
between $n$ and $V$.

A dense gaseous blob that is put in orbit within an ambient gaseous
medium will experience ram pressure from the ambient medium.  If this
ram pressure is stronger than the gravitational restoring force
provided by the potential well in which the dense gaseous blob
resides, then the dense gas will be stripped away.  The critical
condition, when ram pressure equals the gravitational restoring force,
can be determined via
\noindent
\be
\rho V^2 = \frac{GM_\ast M_g}{4\pi R^4},
\ee
\noindent
where $M_\ast$ is the total stellar mass of the GC, $M_g$ and $R$ are
the mass and radius of gas within the GC, $V$ is the orbital velocity
of the GC, and $\rho$ is the density of the ambient medium through
which the GC moves.  We will assume that $M_g=fM_\ast$ and take $R$ to
be the half-mass radius of the cluster.  This yields the following
expression for $V$:
\noindent
\be
\label{e:pram}
V = 580 \,\bigg(\frac{n}{{\rm
    cm}^{-3}}\bigg)^{-1/2}\bigg(\frac{f}{0.1}\bigg)^{1/2}\bigg(\frac{R}{3\pc}\bigg)^{-2}\bigg(\frac{M}{10^5\Msun}\bigg)\, \kms.
\ee
\noindent
Notice the relatively strong dependence on radius.  Observationally,
the half-light radii and masses of MW GCs are only weakly correlated.
It is possible that any primordial mass-radius relation has been
erased by subsequent dynamical evolution.  Nonetheless, in the absence
of any convincing evidence to the contrary, we assume herein that {\it
  initial} GC masses and radii are uncorrelated.  Unless stated
otherwise, we will assume GC half-mass radii of 3 pc, which is the
average half-mass radius of GCs in the MW as determined from the MW GC
catalog of \citet{Harris96}.  GCs in other galaxies also have average
half-light radii of 3 pc, independent of GC luminosity
\citep[e.g.,][]{Masters10b}.

We emphasize that the relations derived above are only approximate.
Variations by factors of several can occur depending on the detailed
orbital properties, density profile of the GC and ambient ISM, and
equation of state of the ambient ISM.  These relations, and the
discussion to follow, should be interpreted as a guide to the trends
and order of magnitude effects to be expected.

\begin{figure}[!t]
\begin{center}
\vspace{.5cm}
\resizebox{3.3in}{!}{\includegraphics{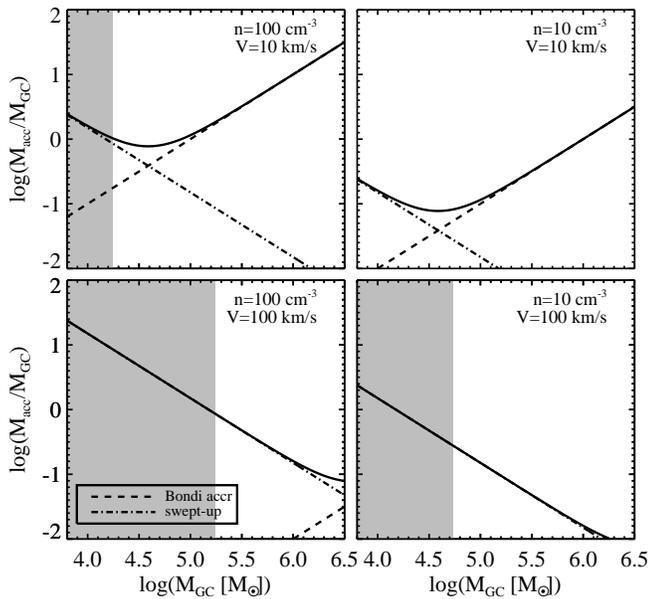}}
\end{center}
\vspace{0.5cm}
\caption{Comparison of the importance of ram pressure, Bondi
  accretion, and ambient ISM sweeping to the development of the gas
  reservoir within GCs.  This figure is similar to Figures
  \ref{fig:main} and \ref{fig:mass}, except that here we show the
  fraction of mass accreted after $10^8$ yr as a function of GC mass
  for a variety of environments.  Shaded zones are regions of
  parameter space where ram pressure is sufficient to remove the GC
  gas supply.  Solid lines show the combined effects of Bondi
  accretion and ISM sweeping.}
\label{fig:facc}
\vspace{0.5cm}
\end{figure}

The relative importance of these three processes is explored in
Figures \ref{fig:main}, \ref{fig:mass}, and \ref{fig:facc}.  In Figure
\ref{fig:main} we compare the importance of ram pressure, Bondi
accretion, and ISM sweeping as a function of the velocity, $V$, and
ambient ISM density, $n$.  Recall that for Bondi accretion $V$ can be
interpreted as the sum in quadrature of the GC gas sound speed and the
relative motion of the GC through the ambient ISM; for other
mechanisms $V$ should be interpreted simply as the relative velocity
between the GC and the ambient medium.  For the ram pressure
calculations, we have assumed that the mass of the GC gas reservoir is
10\% of the total GC mass (i.e., $f=0.1$; this assumption is discussed
at the end of the section).  We have also assumed a GC radius of 3 pc,
as per the discussion above.  Results in Figure \ref{fig:main} are
shown for GC masses of $10^{4.5}\Msun$ and $10^{5.5}\Msun$.  The lower
mass corresponds to the lowest MW GC mass in which multiple
populations have been found \citep{Carretta10b}.  Also, in Figure
\ref{fig:main} we show the relation between $n$ and $V$ for a
gas-dominated, self-gravitating system, for two gas masses labeled in
the figure.  This relation is included as a rough guide to the
properties of the host systems in which GCs may have been found at
their formation epoch.

In Figure \ref{fig:mass} we compare the same processes now as a
function of $V$ and GC mass, for a fixed number density of $n=1\cm$
and $n=100\cm$.  Finally, in Figure \ref{fig:facc} we show the ratio
of total accreted mass to GC mass as a function of GC mass at four
different environments.

There are several important conclusions to be drawn from these
figures.  First, for the {\it present} MW circular velocity
($V\approx220\kms$) ram pressure will be sufficient to remove the gas
within the GC when the GC crosses the MW disk (where $n\sim1\cm$) for
GC masses $\lesssim10^{4.5}\Msun$.  The circular velocity of the LMC
is a factor of $\approx3$ lower than the MW \citep{Kim98}, and so ram
pressure will be able to remove the GC gas supply only in clusters a
factor of $\approx3$ less massive\footnote{Notice that ram pressure
  scales as $V^2$ and the restoring force scales as $M_\ast M_g\propto
  M_\ast^2$.  The critical GC mass therefore depends linearly on the
  velocity.}, or for $M<10^4\Msun$.  We therefore expect that {\it
  young} clusters forming with masses $\lesssim10^{4.5}\Msun$ in the
MW should be truly coeval, while young clusters of comparable mass
forming in the LMC may host multiple populations.  The critical mass
for MW GCs is considerably higher than typically assumed because of
the high GC gas mass fraction assumed herein (10\%), which is
appropriate when considering the early development of GCs, when
mass-loss rates are high.  When one considers ram pressure effects in
present day GCs, the gas mass fractions considered are typically less
than 1\%, and so ram pressure is much more effective
\citep[e.g.,][]{Frank76, Gnedin02}.

Of course, the ancient MW GCs almost certainly formed in environments
very different from the present day MW.  While the present MW velocity
is $V\approx220\kms$, the velocity of the typical progenitor system in
which the MW GCs formed was likely at least a factor of 10 lower.
Thus, less massive ancient GCs may have also survived the effects of
ram pressure (if we assume $V_{\rm proj}\approx20\kms$ and $n=1\cm$
the critical mass becomes $10^{3.5}\Msun$).  It is because of this
fact that we emphasized that the critical mass of $\sim10^{4.5}\Msun$
should be manifest in clusters forming in the MW and LMC at the {\it
  present epoch}.  Without knowing in detail the properties of the
progenitor systems in which MW GCs formed, one cannot discuss
quantitatively the effects of ram pressure on the early development of
the ancient MW GCs.

As mentioned in $\S$\ref{s:intro}, the only class of star clusters
known {\it not} to contain multiple stellar populations are the open
clusters in the MW \citep{deSilva09, Martell09}.  This population is
much less massive than either the MW GCs or the LMC clusters, with
typical masses of $\sim10^3\Msun$.  Based on Figure \ref{fig:mass}, it
is not surprising that such low mass systems do not contain multiple
populations because ram pressure would be effective at stripping gas
within the GC for any plausible formation environment.

The right panel of Figure \ref{fig:mass} shows the effects of ram
pressure, Bondi accretion, and ISM sweeping when the ambient ISM
density is $n=100\cm$.  These densities are very high compared to
typical ISM environments in the local Universe, but may be common at
the epoch of massive GC formation.  For example, the bulge of the MW
likely formed rapidly at $z\gtrsim2$ \citep{Zoccali06, Ballero07}, has
a stellar mass of $10^{10}\Msun$ and a half-mass radius of
$\approx1\kpc$ \citep{Binney97}.  If all the bulge stars were
converted into gas, the gas density would be $n\sim100\cm$ at the
epoch of GC formation.

Finally, Figure \ref{fig:facc} shows the total mass accreted within
$10^8$ yrs divided by the initial GC mass as a function of GC mass.
Depending on the formation environment, clear trends are predicted for
the total accreted mass as a function of initial GC mass.  In
particular, if the relative velocities were low and the ambient
densities high (upper left panel) at the epoch of formation, one would
expect an increasing fraction of accreted material as GC mass
increases.

In this section we have assumed that the gas mass within the GC is
only $f=10\%$ of the GCs stellar mass.  Of course, by the time the GC
forms its second generation of stars, the gas mass fraction must be of
order unity (or significantly greater than unity, depending on the
star formation efficiency) in order to satisfy the observational
constraint that the first and second generations are approximately
equal in mass.  Increasing $f$ to unity results in a stronger
restoring force and therefore a reduced ability of ram pressure to
remove gas within the GC.  The critical zone for ram pressure
stripping is therefore time-dependent.  A more thorough investigation
that what has been presented herein will therefore require
time-dependent model calculations.  In particular, the initial
development of a gaseous reservoir will depend on the detailed
hydrodynamical interaction between the ambient medium and AGB outflows
in the presence of the GC potential well.

\section{Constraints from intermediate-age clusters in the LMC}

In this section we present new evidence for an LMC cluster mass
threshold below which LMC clusters appear to be truly coeval.  This
result is then discussed in the context of early GC evolutionary
scenarios.

Multiple stellar populations have been detected in intermediate-age
($\sim1$ Gyr) and old LMC clusters \citep{Mackey08, Goudfroij09,
  Milone09, Mucciarelli09}.  The old clusters have masses of
$\approx10^{5.5}\Msun$ \citep{Mackey03a}; the masses of the
intermediate-age clusters will be derived below.  Intermediate-age
clusters possess the property that age differences of $\sim10^8$ yr
are readily observable in the main sequence turn-off point because at
these ages the turn-off point is a strong function of time.

The identification of multiple populations in intermediate-age
clusters is important for several reasons.  First, the birth
environment of such clusters cannot be dramatically different from the
present day LMC, in contrast with the old clusters.  This opens the
possibility of making more direct links between the clusters and their
environment than is possible with the old clusters.  Second, as
discussed in $\S$\ref{s:dm}, these observations argue strongly against
GC formation at the center of their own dark matter halos as a means
to produce multiple populations.

\begin{figure}[!t]
\resizebox{3.6in}{!}{\includegraphics{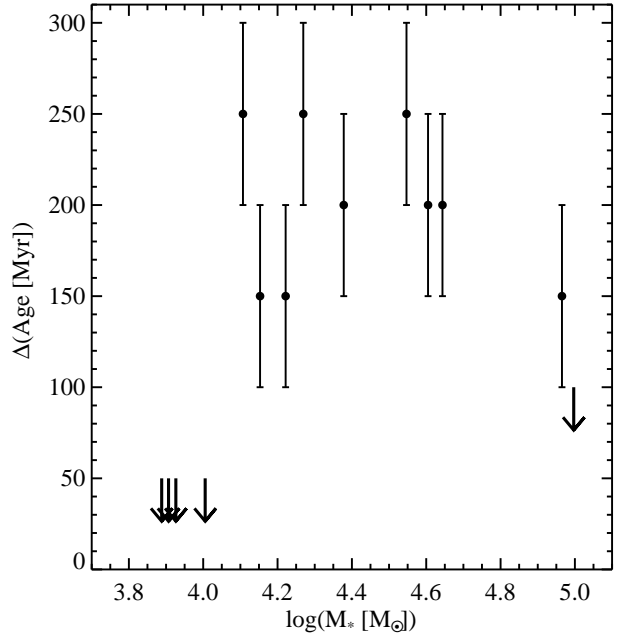}}
\caption{Internal age spread in intermediate-age LMC clusters as a
  function of cluster stellar mass.  The age spreads are adopted from
  \citet{Milone09} and are based on isochrone fits to {\it HST}-based
  CMDs.  Stellar masses are derived herein.  Upper limits indicate
  that the cluster is consistent with a single-age population.  Notice
  that below log$(M_\ast/\Msun)\approx4.0$ all LMC clusters are
  consistent with being coeval.  The most massive cluster in this
  sample, NGC 1978, is also consistent with being coeval, although
  this cluster has a variety of peculiar properties; see the text for
  details.  The lowest mass cluster has been shifted slightly in mass
  for clarity.}
\label{fig:lmc}
\vspace{0.5cm}
\end{figure}

\citet{Milone09} has recently analyzed archival {\it HST} images of 16
intermediate-age clusters in the LMC.  These authors derive not only
average ages, metallicities, distances, and reddening values, but also
the internal spread in age for each cluster in their sample.  Only
five of the clusters in their sample were consistent with being
coeval.  We are interested in knowing whether the internal age spreads
measured by Milone et al. correlate with the mass of the cluster.  We
have therefore derived stellar masses for the clusters in Milone et
al., in the following way.  The average age of each cluster is
converted into a $V-$band mass-to-light ratio, $M/L_V$, via the
stellar population synthesis models of \citet{Conroy09a} as updated in
\citet{Conroy10c}.  We assume $Z=0.006$ for the young clusters in the
LMC and a \citet{Kroupa01} IMF.  We then adopt $V-$band magnitudes for
the Milone et al. clusters from the data compilation of
\citet{vandenBergh81} (there are fourteen clusters in common).  These
photometry are then corrected for extinction and converted to absolute
luminosities via the $E(B-V)$ values and distance moduli listed in
Milone et al.  Finally, stellar masses are estimated by combining the
expected $M/L_V$ and absolute magnitudes.  The majority of the
clusters have similar ages and so the estimated $M/L_V$s do not vary
considerably across the sample.  The following result therefore does
not depend critically on the adopted modeling approach.

In Figure \ref{fig:lmc} we show the relation between the total stellar
mass and internal age spread for the clusters in \citet{Milone09}.  It
is remarkable that all clusters with masses $\lesssim10^4\Msun$ show
no evidence for multiple populations.  In the LMC, there appears to be
a critical mass below which the formation of multiple populations is
suppressed.  This critical mass is fully consistent with our estimates
of the effects of ram pressure effects in the LMC, where for $n=1\cm$
and $V=70\kms$ we find a critical mass of $M\approx10^4 \Msun$.  Given
the sensitivity to parameters such as the GC radius, a more
quantitative comparison must await detailed observations of these
clusters.

The most massive cluster in Figure \ref{fig:lmc}, NGC 1978, also shows
no evidence for multiple populations.  It is worth noting that
\citet{Milone09} extract the CMD for their target clusters from the
cluster core for all clusters except for NGC 1978 because of crowding.
Since the second generation appears to be more spatially concentrated
than the first (see $\S$\ref{s:intro}), this may explain the observed
lack of significant internal age spread in this cluster.  This cluster
has a number of peculiar properties, including a high ellipticity
\citep{Geisler80}, a lack of S stars, and odd carbon and oxygen ratios
in its AGB stars \citep{Lederer09}.  Owing to all of these facts, we
are reluctant to place much weight on the result that this massive
cluster does not show obvious signs of internal age spreads.

\citet{Bastian09} raise the possibility that the observed broadening
of the CMD of intermediate-age LMC clusters is in fact due to stellar
rotation, and not multiple populations.  We believe this scenario to
be implausible for several reasons.  First, our results in Figure
\ref{fig:lmc} strongly suggest that the width of the main sequence
turn-off is related to the total cluster mass.  This correlation is
difficult to explain if the width was due to stellar rotation.
Second, no such broadening is observed in the intermediate-age MW open
clusters, including the Hyades, Pleiades, and Praesepe
\citep[e.g.,][]{Vandenberg84, Griffin88, Pinsonneault04, An07a}.
These MW open clusters have masses of order $10^3\Msun$, and so in our
scenario they should be truly coeval, as observed.  The scenario of
Bastian \& de Mink can be reconciled with the open cluster data and
low mass LMC data only by supposing an {\it ad hoc} correlation
between stellar rotation rate and cluster mass, which seems
implausible.


\section{Summary, open issues and future directions}
\label{s:res}

In this work we have presented a comprehensive model for the early
(several $10^8$ yr) evolution of massive star clusters.  Our model
considers the importance of type II and prompt type Ia SNe, accretion
onto the GC from the ambient ISM, the ability of a GC to retain its
internal gas supply in the face of ram pressure, and the effect of the
Lyman-Werner photon flux density on the ability of the young GC gas to
form molecules and, ultimately, stars.

This model definitively addresses two of the three major issues in
early GC evolution identified in $\S$\ref{s:prev}: how GCs can retain
a gaseous reservoir in the face of ram pressure, and which stars are
responsible for processing material to $T>10^7$K.  From consideration
of the formation environments of both old MW GCs and intermediate-age
LMC clusters, we have shown that ram pressure stripping naturally
explains the observed bifurcation between clusters that do and do not
show evidence for multiple stellar populations.  Based on several
independent arguments, including the similar $\sim10^8$ yr timescale
of many physical mechanisms and the lack of internal spread in Fe and
Ca in most clusters, we strongly favor massive AGB stars as the source
of the processed material.

The final open issue is in understanding how a second generation can
form with a current total mass comparable to the first generation, at
least for old MW GCs where data are available.  We have considered
accretion from the ambient ISM as a viable mechanism to provide
copious amounts of gas to the GC.  However, our solution to this last
issue currently has little direct empirical verification, and may in
fact have trouble reproducing the observed correlations in abundance
ratios, as discussed below.  We now discuss a variety of testable
implications of our proposed solution, and comment briefly on several
open issues.

As a consequence of significant accretion from the ambient ISM, we
expect several trends with GC mass (see e.g., Figure \ref{fig:facc}).
For example, as the fraction of pristine gaseous material increases,
we expect a shorter Na-O correlation, a greater abundance of Li and F,
and in general we expect the stars with anomalous abundances to be
less anomalous.  The sign of the trend with mass depends on the
dominant accretion process, which in turn depends on the formation
environment.  Bondi accretion scales as $M^2$ while the mass provided
by AGB winds scales with $M$ so the pristine fraction increases with
mass.  ISM sweeping scales as $R^2$ which is only weakly correlated
with $M$ at the present epoch, and so if this process is dominant we
would expect the pristine fraction to decrease with mass.

Observations of young and intermediate-age clusters will allow
identification of the relevant mechanism because the formation
environment of young clusters is not so dissimilar from their present
environment.  The old GCs within dwarf spheroidals and dwarf
irregulars may also shed light on this issue.  The GCs within dwarf
galaxies almost certainly formed where they are now observed since the
stellar accretion/merger rate onto dwarfs is expected to be very low
in a $\Lambda$CDM cosmology, and so the present day conditions of the
host dwarf galaxies cannot be very different from the formation
environment of the GCs.

\citet{Carretta06} and \citet{Carretta10c} have investigated the extent
of the Na-O correlation as a function of global parameters and have
found evidence that the correlation is more extended in higher mass
GCs, and for GCs with more extended orbits.  This result is consistent
with ISM sweeping being the dominant accretion process, but we caution
that the data show large scatter, and the present GC mass is poorly
correlated with its mass at formation due to orbit-dependent mass-loss
effects.  In general, we expect that the extent of the Na-O
correlation should depend on the formation environment at fixed GC
mass, since the formation environment determines the amount of
accreted material.  Another effect might be the mass-dependent ability
of a GC to retain the winds from AGB stars.  In any event, it is clear
that such trends hold the promise of isolating which physical process
dominates the accretion of pristine material onto GCs.

If accretion from the ambient ISM was important, then it is somewhat
puzzling why the Fe and Ca abundances are so uniform between the first
and second generations.  The only plausible explanation is that at
early times, during the formation of the ancient MW GCs, the spread in
Fe and Ca abundances within the MW progenitor system was quite small.
A clear prediction of the accretion scenario is that the
intermediate-age LMC clusters showing multiple populations should show
a much larger internal spread in Fe, Ca, etc. abundances, on the order
of the spread in abundances of these elements within the LMC as a
whole.

The MW bulge GC Terzan 5 may provide additional insight.
\citet{Ferraro09} has convincingly shown that this GC has a split
horizontal branch (HB) with the more luminous branch having a much
higher Fe abundance ([Fe/H]$\sim+0.3$) than the less luminous branch
([Fe/H]$\sim-0.2$).  \citet{DAntona10} argue that the split HB is
consistent with an internal age spread of several $10^8$ yr and
enhanced He in the metal-enriched branch.  Since this cluster is in
the MW bulge and Fe-rich, it probably formed in the bulge.  D'Antona
et al. suggest that the second stellar generation may have acquired
its high Fe abundance via accretion from the ambient, Fe-rich ISM.
Self-enrichment is unlikely because the required Fe mass to produce
the observed high Fe abundance would require so many type II SNe that
the cluster would easily become unbound, unless Terzan 5 was
significantly more massive in the past.  Terzan 5 may therefore be an
ideal cluster to look for further evidence for the importance of
ambient ISM accretion in the formation of multiple stellar
populations, e.g., by observing the Li and F abundances in this
cluster.  If the proposed scenario for Terzan 5 is generic then we
might expect to find significantly separated Fe abundances within the
other bulge metal-rich GCs as well.

In the present work we have not attempted to make specific predictions
for the extent of correlations amongst various elemental abundances.
Such predictions would provide a powerful constraint on the model.
Qualitative considerations suggest that substantial accretion from
the ISM may yield a very short Na-O anti-correlation, owing to the
ISM having abundance ratios similar to the first generation stars.
Unfortunately, uncertainties in the AGB yields greatly complicate any
attempted comparison to observed abundances.  A fruitful avenue for
future work will consist of a detailed comparison between predicted
and observed abundance correlations in light of the uncertain AGB
yields \citep[see][for an initial attempt in this
direction]{DErcole10b}.

The most massive GCs, including $\omega$Cen, NGC 2808, M22, and M54,
deserve special mention.  Each of these clusters show multiple, {\it
  distinct} sequences in the CMD, and the former two are unique in
that they display multiple main sequences.  The CMD morphology of
these clusters suggests very high He abundances in the second (and
third) stellar generations.  The discrete sequences and high He
abundances argue against significant dilution from ambient ISM
accretion.  This can be accommodated in our model if ISM sweeping is
the dominant accretion mechanism because this mechanism becomes
increasingly less important at high GC masses (see the bottom panels
of Figure \ref{fig:facc}).  Although we are then left with the
original problem of explaining how so many second generation stars
could form from the mass lost by the first generation.  We speculate
that perhaps the most massive GCs formed at the centers of their own
dark matter halos, and contained even more stars in their past.  The
precursors of these systems could be nuclear star clusters, which are
common in low-mass galaxies.  Regardless of these details, we stress
caution when attempting to incorporate the massive GCs into any
framework for the early evolution of GCs since many of their
observational characteristics differ qualitatively from the lower mass
GCs.

An exciting direction for future observational work is characterizing
the incidence of multiple populations within the numerous young and
intermediate-age clusters in M31.  These clusters span a range in mass
from $10^3\Msun$ to $10^5\Msun$ \citep{Caldwell09}.  Based on ram
pressure arguments, we expect the higher mass clusters to show
evidence for multiple populations, but not the lower mass clusters.
Assuming $n=1\cm$ and $V=250\kms$ for disk of M31 \citep{Widrow03}, we
predict that the critical mass will be $\approx10^{4.5}\Msun$.
However, as stressed in previous sections, the relevant velocity is
the relative velocity between the ambient ISM and the GC.  Since many
of the young M31 GCs are orbiting within the disk of M31, their
relative velocity will be considerably less than the circular
velocity, and so the critical mass may be considerably lower.
Confrontation of our model predictions with the properties of the M31
GCs must take these details into account.

Since these M31 clusters have not experienced significant dynamical
evolution, we may expect stronger correlations between the extent of
any abundance anomalies and cluster mass.  For the youngest clusters,
with ages of several $10^8$ yr, we might even hope to {\it directly
  observe} the formation of the second generation.  Such an
observation would provide definitive proof that GCs need not form
within dark matter halos to produce multiple stellar generations, and
would also demonstrate that AGB stars are the polluters, since the AGB
polluter scenario is expected to produce a second generation on a
timescale of several $10^8$ yr.

In the past several years it has become clear that star clusters, once
thought to be simple systems, in fact show an internal complexity that
increases with increasing mass.  Observations indicate that the lowest
mass systems, i.e., the open clusters, appear to be truly coeval and
mono-metallic.  At higher masses one observes age spreads of several
$10^8$ yr and internal variation in the light elements.  At still
higher masses ($\gtrsim10^6\Msun$) it appears that the cluster
potential well is deep enough to retain SNe ejecta and hence
self-enrich, perhaps because these massive clusters form at the center
of their own dark matter halos.  This last phenomenon is observed not
only in the most massive MW GCs, but also appears to be present in the
most massive GCs within other galaxies as well \citep{Strader08,
  Bailin09}.  At the very least, it is now abundantly clear that star
clusters are not simple systems.


\acknowledgments 

We acknowledge fruitful conversations with Bruce Draine, and thank
Alvio Renzini for comments on an earlier draft.  The referee is
thanked for comments that improved the quality of the manuscrit.  This
work made extensive use of the NASA Astrophysics Data System and of
the {\tt astro-ph} preprint archive at {\tt arXiv.org}.


\end{document}